\documentclass[aps,pre,twocolumn,superscriptaddress,floatfix]{revtex4}
\usepackage{graphicx}
\usepackage[dvips,unicode,colorlinks,linkcolor=blue,citecolor=blue,urlcolor=blue]{hyperref}

\begin{document}

\title{Nonlinear Breathing-like Localized Modes in C$_{60}$ Nanocrystals}

%

\author{Alexander V. Savin}

\affiliation{Semenov Institute of Chemical Physics, Russian Academy
of Sciences, Moscow 119991, Russia}
\affiliation{Nonlinear Physics Center, Research School of Physics and
Engineering, Australian National University, Canberra ACT 0200,
Australia}

\author{Yuri S. Kivshar}
\affiliation{Nonlinear Physics Center, Research School of Physics and
Engineering, Australian National University, Canberra ACT 0200,
Australia}

\begin{abstract}
We study the dynamics of nanocrystals composed of C$_{60}$ fullerene molecules.
We demonstrate that such structures can support long-lived strongly localized nonlinear
oscillatory modes, which resemble discrete breathers in simple lattices. We reveal
that at room temperatures the lifetime of such nonlinear localized modes may exceed 
tens of picoseconds; this suggests that C$_{60}$ nanoclusters should demonstrate
anomalously slow thermal relaxation when the temperature gradient decays in accord 
to a power law, thus violating the Cattaneo-Vernotte law of thermal conductivity.
\end{abstract}
\maketitle

\section{Introduction}

The fullerene era was started in 1985 with the discovery of a stable C$_{60}$ cluster
and its interpretation as a cage structure with a soccer ball shape~\cite{p2}. By now,
many structural, electronic, and vibrational properties of fullerenes have been studied 
in detail.
In spite of the rapidly increasing interest in new forms of fullerenes, icosahedral C$_{60}$,
the "most beautiful molecule" \cite{p3}, remains at the focus of active research as a
prototype fullerene system. Fullerene molecule C$_{60}$ keeps a prominent position
primarily for the high symmetry which greatly facilitates theoretical computations and the
interpretation of experimental data, but also because it is only fullerene which can be grown
in the form of large and nearly perfect single crystals. At room temperature, the diffraction
experiments~\cite{ref1} show that the  C$_{60}$ molecules are centered on sites of 
a face-centered-cubic (fcc) Bravais lattice. Each C$_{60}$ molecule in the solid-state phase
is coupled to other C$_{60}$ molecules through weak Van-der Waals forces. 

By now, the internal oscillations of the fullerene molecule C$_{60}$ are well studied, for
the case of an isolated molecule and also for C$_{60}$ crystalline structures. Various 
theoretical and numerical methods have been developed and employed. However, the dynamics 
of fullerenes in the nonlinear regime and a role of nonlinear localized modes in such structures
remain largely unexplored.

The main aim of this paper is to study strongly localized nonlinear oscillatory modes in such
nanocrystals which resemble the discrete breathers of much simpler nonlinear lattices.

High symmetry of C$_{60}$ molecules allows the existence of internal nonlinear oscillations
which do not alter substantially the spherical shape of the fullerene molecule. Such oscillations
are excited by rotational degrees of freedom of pentagons and hexagons of the structure, with
a variation of valent bonds. Resonant coupling of such high-amplitude oscillations with 
the neighboring C$_{60}$ molecules will remain weak, because due to conservation of a spherical
shape the molecules will interact as large particle through Lennard-Jones potentials.  
This observation suggests that such structures can provide an ideal environment for the 
existence of nonlinear modes strongly localized in the space~\cite{r1,r2}.

Here we demonstrate that the C$_{60}$ crystals can support long-lived strongly localized 
nonlinear modes which closely resemble the so-called discrete breathers in simple nonlinear
lattices, and affect dramatically the heat relaxation of composite fullerene structures 
that obeys a power low instead of an exponential law.

\section{Model and simulation methods}

\begin{figure*}[tb]
\begin{center}
\includegraphics[angle=0, width=1\textwidth]{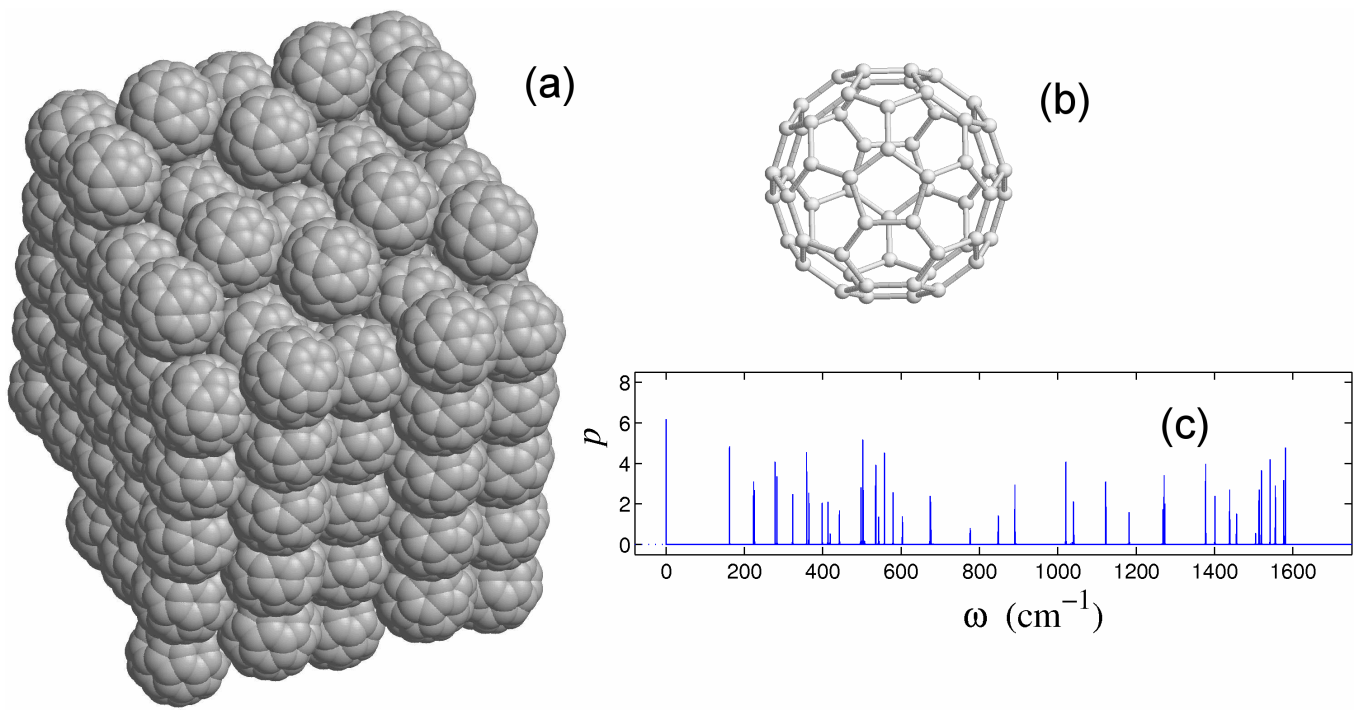}
\end{center}
\caption{\label{fg01}\protect
 (a) General view of a cubic nanocluster composed of $5\times 5\times 5$
fullerene molecules C$_{60}$ (buckyballs C60-Ih), (b) isolated molecule  C$_{60}$
and (c) dimensionless density of its frequency spectrum for thermal vibrations at $T=3$K.
}
\end{figure*}

To analyze the dynamics of of localized nonlinear modes in three-dimensional C$_{60}$ crystals,
we consider a cubic nanocluster of C$_{60}$  molecules, as shown in Fig.~\ref{fg01}~(a).

Inside of each C$_{60}$ molecule of the nanocluster two neighboring carbon atoms create 
a valent bond. We describe the energy associated with deformation of the valent bond
created by two carbon atoms by the following interaction potential
\begin{equation}
U_1(\rho)=\epsilon_1\{\exp[-\alpha_0(\rho-\rho_0)]-1\}^2,
\label{f1}
\end{equation}
where $\rho$ is the length of the valent bond, $\epsilon_1=4.9632$~eV and $\rho_0=1.418$~\AA~
is the energy and the equilibrium length of the bond, parameter $\alpha_0=1.7889$~\AA$^{-1}$.
Each carbon atom is placed at the vertex of three planar valent angles.
The corresponding deformation energy of a plane valent angle created by three atoms
can be described by the interaction potential of the form,
\begin{equation}
U_2(\varphi)=\epsilon_2(\cos\varphi+1/2)^2,
\label{f2}
\end{equation}
where $\varphi$ is value of the valent angle (equilibrium value of the angle $\varphi_0=2\pi/3$),
energy $\epsilon_2=1.3143$~eV. Each valent bond is simultaneously belong to two planes, so
that the deformation energy of the angle formed by two such planes can be described
by the following interaction potential,
\begin{equation}
U_i(\phi)=\epsilon_i(1-\cos\phi),
\label{f3}
\end{equation}
where $\phi$ --is the corresponding angle (in equilibrium, $\phi=0$), and the index
$i=3$, 4, 5 describe the type of the valent angle.
Energy $\epsilon_3=\epsilon_4=0.499$~eV, $\epsilon_5\ll \epsilon_4$, so that
the latter contribution to the total energy can be neglected.

More detailed discussion and motivation of our choice of the interaction potentials
$U_1$,...,$U_4$ can be found in Ref.~\cite{skh10}. Such potentials have been employed
for modeling of thermal conductivity of carbon nanotubes~\cite{skh09,shk09} 
graphene nanoribbons~\cite{skh10,sk11}
and also in the analysis of their oscillatory modes~\cite{sk09,sk10,sk10prb}.

We describe the interaction of carbon atoms belonging to different
C$_{60}$ molecules by the Lennard-Jones 12-6 potential
\begin{equation}
U_{LJ}(r)=4\epsilon_{LJ}[(\sigma/r)^{12}-(\sigma/r)^6],
\label{f4}
\end{equation}
where $r$ -- distance between the atom centers, energy $\epsilon_{LJ}=0.004556$~eV,
and parameter $\sigma=3.851$~\AA.

To verify our results, we also employ the simpler model where all interacting potentials between
the atoms are described, instead of the set of the potentials $U_1$,..., and $U_4$,
by the Brenner potential~\cite{br1}.
\begin{table*}[tb]
\centering\noindent
\caption{
Values $\omega$ (in cm$^{-1}$) for the 46 distinct normal mode frequencies of an isolated
C$_{60}$ molecule ($n$ is mode degeneracy). Notations of the modes are from Ref.~\cite{p1}.}
\label{tab1}
\begin{tabular}{lcl|ccl|ccl|ccl}
\hline
\hline
~mode& ~~~~~$\omega$~~~~~ & n~~~ & mode & ~~~~~$\omega$~~~~~
&   n~~~ & mode & ~~~~~$\omega$~~~~~  &   n~~~ & mode & ~~~~~$\omega$~~~~~  &   n~~~\\
\hline
H$_g$(1)  & 162.1 & 5~& H$_g$(3)   & 497.4 & 5~& A$_u$      & 776.0  & 1~& G$_g$(5) & 1437.5 & 4\\
T$_{3u}$(1)& 223.2& 3 & T$_{3u}$(2)& 497.8 & 3 & G$_u$(4)   & 847.7  & 4 & T$_{1g}$(3) & 1454.9 & 3\\
G$_u$(1)  & 225.5 & 4 & H$_u$(4)   & 501.8 & 5 & T$_{3u}$(3)& 890.2  & 3 & A$_g$(2) & 1503.2 & 1\\
H$_u$(1)  & 278.2 & 5 & T$_{1u}$(2)& 503.6 & 3 & H$_g$(5)   & 1020.0 & 5 & T$_{3u}$(5) & 1511.7 & 3\\
H$_g$(2)  & 282.8 & 5 & G$_g$(3)   & 534.7 & 4 & G$_g$(4)   & 1039.8 & 4 & T$_{1u}$(4) & 1512.9 & 3\\
G$_g$(1)  & 323.1 & 4 & G$_u$(2)   & 535.0 & 4 & H$_u$(5)   & 1121.8 & 5 & H$_g$(7) & 1518.2 & 5\\
H$_u$(2)  & 358.6 & 5 & T$_{3g}$(2)& 542.5 & 3 & T$_{1u}$(3)& 1181.2 & 3 & G$_u$(6) & 1540.6 & 4\\
T$_{1u}$(1)& 364.3& 3 & H$_u$(3)   & 557.1 & 5 & T$_{3g}$(4)& 1267.6 & 3 & G$_g$(6) & 1553.8 & 4\\
T$_{1g}$(1)& 398.5& 3 & G$_u$(3)   & 579.3 & 4 & H$_g$(6)   & 1270.3 & 5 & H$_u$(7) & 1575.0 & 5\\
G$_g$(2)  & 413.1 & 4 & T$_{3g}$(3)& 603.1 & 3 & T$_{3u}$(4)& 1272.0 & 3 & H$_g$(8) & 1579.3 & 5\\
A$_g$(1)  & 418.6 & 1 & H$_g$(4)   & 673.7 & 5 & H$_u$(6)   & 1376.2 & 5 & ~ & ~ & ~\\
T$_{3g}$(1)& 442.3& 3 & T$_{1g}$(2)& 675.7 & 3 & G$_u$(5)   & 1399.8 & 4 & ~ & ~ & ~\\
\hline
\hline
\end{tabular}
\end{table*}

\section{Intramolecular modes}

The high symmetry of an isolated, isotopically pure C$_{60}$ molecule in its equilibrium
configuration imposes very strong constraints on the form of the normal mode displacement patterns.
All 60 sites are symmetry-equivalent, and the point group is the full icosahedral group $I_h$.
Although there are 180 degrees of freedom (3$\times$60) for each C$_{60}$ molecule, the icosahedral
symmetry gives rise to a number of degenerate modes, so that only 46 distinct mode frequencies
are expected for the C$_{60}$ molecule. All vibrational modes of C$_{60}$ are distributed \cite{p4,p1}
as 2A$_g$+A$_u$+4T$_{1g}$+5T$_{1u}$+4T$_{3g}$+5T$_{3u}$+6G$_g$+6G$_u$+8H$_g$+7H$_u$.

Dynamics of an isolated C$_{60}$ fullerene molecule is described by a system of $3\times 60$
nonlinear equations of the standard form,
$M\ddot{\bf u}=-\partial H/\partial{\bf u}$,
where $M=12\cdot1.6603\cdot 10^{-27}$ kg -- mass of the carbon atom,
system Hamiltonian is $H=M(\dot{\bf u},\dot{\bf u})/2+E({\bf u})$,
and 180-dimensional vector ${\bf u}=\{(x_n,y_n,z_n)\}_{n=1}^{60}$
defines the coordinates of all atoms of the complex molecule.
Potential energy of the molecule $E({\bf u})$ is composed by all
types of interactions described by the potentials $U_1$,...,$U_4$ discussed above.

To find the equilibrium stationary state of the  C$_{60}$ molecule, we need to solve
numerically the minimization problem for the energy functional, $E({\bf u})\longrightarrow\min$.
In the model under consideration, the molecule C$_{60}$  is described by the set of
coordinates ${\bf u}_0$, and it has a shape of perfect icosahedral [see Fig.~\ref{fg01}~(b)]
with the radius $R=3.514$~\AA, for which all valent bonds have the equilibrium length
$\rho=\rho_0$.

 Linearizing the model near the equilibrium stationary state, we obtain a system f linear equations
$M\ddot{\bf u}={\bf A}{\bf u}$, where ${\bf A}$ is  $180\times 180$ dimensional matrix,
${\bf A}=\{\partial E/\partial u_i\partial u_j|_{{\bf u}={\bf u}_0}\}_{i=1,j=1}^{180,180}$.
Diagonalizing  the matrix ${\bf A}$ we find all 180 eigenmodes of the molecule.
The first six zero-frequency modes correspond to rigid translations and rotations of the
molecule. Other 174 modes are generated and, in view of their degeneracy,
there are exists 46 distinct modes -- see Table~\ref{tab1}.

The frequencies of all oscillatory eigenmodes of an isolated C$_{60}$ molecule are
summarized in Table~\ref{tab1}. All frequency are located in the interval 
$162.1\le\omega\le 1579.3$~cm$^{-1}$.
Definitely, the model with only 6 parameters is not able to predict exactly the frequencies
obtained in the framework of the first principles calculations or experimental results, see~\cite{p1}.
Nevertheless, this model demonstrates a correct structure of the eigenmode frequencies
and predict rather well the frequency band measured experimentally,
$272\le\omega\le 1575$~cm$^{-1}$.

It is important to mention that the use of the Brenner potentials~\cite{br1} instead of the
set of the potentials $U_1$,...,$U_4$ leads much broader frequency band,
$182.9\le\omega\le 1738.9$~cm$^{-1}$. In addition, the Brenner potential do not predict
correctly the degeneracy of the frequency spectrum, so that the equilibrium shape  of the
C$_{60}$ molecule deviates from the icosahedra shape.

In order to model thermal oscillations of the molecule, we start from its equilibrium state
and excite each of its eigenmode with the energy $k_BT$. The frequency spectrum of the C$_{60}$
molecule at the temperature $T=3$K is shown in the Fig.~\ref{fg01}~(c). As expected, 
the spectrum is discrete,
and the number of peaks coincides with the number of eigenmodes, whereas the peak positions
coincide with the corresponding eigenfrequencies.  At low temperatures, the oscillations are
linear, so that the spectrum corresponds to a system of uncoupled linear oscillators.

If we increase the temperature to $T=30$K, the discrete nature of the spectrum remains unchanged,
but high-frequency modes demonstrate some energy exchange and interaction. For much higher
temperatures such as $T=300$K the discrete nature of the spectrum is lost and the 
oscillations become strongly nonlinear demonstrating interaction with strong energy exchange.

\section{Localized nonlinear modes}

The C$_{60}$ molecules occupy fcc sites in the solid phase, and at room temperature,
the rapid rotational diffusion of each molecule leads to an effective fcc crystal structure.
To study localized modes of the fullerene molecule placed in a crystal, we consider a cubic
nanocluster composed of $5\times 5\times 5$ molecules C$_{60}$, as shown in Fig.~\ref{fg01}~(a).

First, we find the ground state of the nanocluster by minimizing its energy. Next, for
the central molecule we excite its $n$-th eigenmode ($n=7,...,180$) with the energy $E_0$
and study the decay of the kinetic energy in the cluster. Due to the nonvalent coupling 
between the fullerenes, the energy is transferred from one molecule to another.
To model the energy spreading in an effectively infinite crystal, we introduce lossy boundary
conditions, when all surface molecules experience damping with the relaxation time $t_r=20$~ps.
If the excitation of the certain eigenmode lead to the creation of a localized state, or breathing-like
mode, the kinetic energy will not vanish but instead will approach a certain nonzero value.
Otherwise, the initial excitation will vanish completely, so that the
kinetic energy of the central molecule $E_k\rightarrow 0$ when $t\rightarrow\infty$.
\begin{figure}[tb]
\begin{center}
\includegraphics[angle=0, width=1\linewidth]{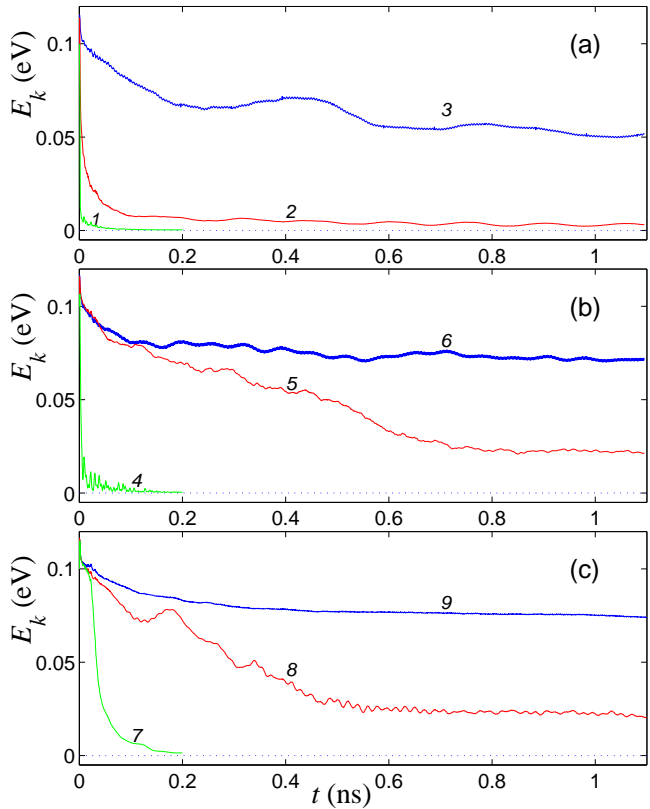}
\end{center}
\caption{\label{fg02}\protect
Evolution of the kinetic energy $E_k$ of the central fullerene
 C$_{60}$ molecule in nanocluster composed of $5\times 5\times 5$
 molecules packed in the form of a fcc cubic crystal. At the initial
 moment $t=0$ only one eigenmode is excited for the central molecule with the energy
$E_0=0.2$eV. (a) curve 1 show the decay of the mode H$_g$(1)
(frequency $\omega=162.2$ cm$^{-1}$), curve 2 -- the mode H$_u$(4) ($\omega=501.8$~cm$^{-1}$),
and the curve 3 -- the mode T$_{3g}$(1)($\omega=442.3$~cm$^{-1}$);
(b) curve 4 gives the dependence for the mode A$_g$(1)($\omega=418.6$~cm$^{-1}$),
curve 5 -- the mode T$_{3g}$(4) ($\omega=1267.6$~ cm$^{-1}$),
curve 6 -- the mode T$_{1g}$(2) ($\omega=675.7$~cm$^{-1}$);
(c) curve 7 describes the decay of the mode A$_u$ (frequency $\omega=776.0$~cm$^{-1}$),
curve 8 -- the mode G$_g$(5) ($\omega=1437.5$~cm$^{-1}$),
and the curve 9 -- the mode T$_{1g}$(3) ($\omega=1454.9$~cm$^{-1}$).
}
\end{figure}

First, we excite with the energy $E_0=0.2$ the $n$-th mode ($n=7,...,180$) of the central molecule
of the fullerene crystal. Depending on the type of the modes, our numerical results reveal
three different scenarios of the mode relaxation, as shown in Fig.~\ref{fg02} (a-c).
The first scenario is the rapid energy relaxation and decay of the mode oscillations
(see curves 1, 4, and 7); the second scenario is slow relaxation (curves 2, 5, and 8), 
and the third scenario is the generation of stationary spatially localized nonlinear modes
(curves 3, 6, and 9).

\begin{figure}[tb]
\begin{center}
\includegraphics[angle=0, width=1\linewidth]{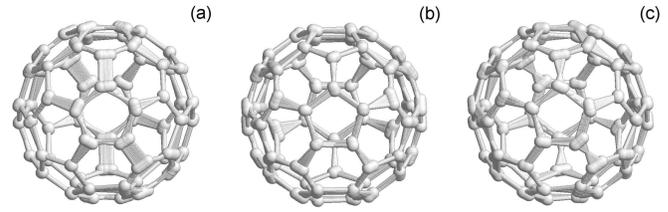}
\end{center}
\caption{\label{fg03}\protect
Selected eigenmodes of an isolated C$_{60}$ molecule which lead
to the generation of spatially localized nonlinear modes (breathers) of the  fullerene crystal.
(a) mode T$_{3g}$(1) (frequency $\omega=442.3$~cm$^{-1}$);
(b) mode T$_{1g}$(2) (frequency $\omega=675.7$~cm$^{-1}$);
(c) mode T$_{1g}$(3) (frequency $\omega=1454.9$~cm$^{-1}$).
}
\end{figure}

The first and second scenarios are observed for the modes associated with a periodic change
of the spherical shape or the size of the C$_{60}$ molecule. The most interesting third 
scenario is
observed for the modes associated with the motion of atoms along the spherical surface so that
the molecule keeps its form almost unchanged. These are the eigenmodes
T$_{3g}$(1), T$_{1g}$(2), and T$_{1g}$(3) presented in Fig.~\ref{fg03}.  For all those modes,
the molecule conserves its spherical shape and therefore the resonant interatomic 
interaction is reduced.

\begin{figure}[tb]
\begin{center}
\includegraphics[angle=0, width=1\linewidth]{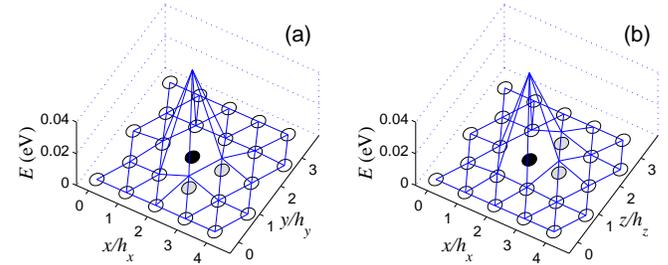}
\end{center}
\caption{\label{fg04}\protect
Spatial profile of the kinetic energy $E$ of the breathing-like nonlinear
localized mode of the central fullerene molecule in the nanocluster of
$5\times 5\times 5$ of C$_{60}$ molecules in the planes (a) $xy$ and (b) $xz$,
intersecting the central molecule (shown by a black circle)
For the generation of such breathers, the eigenmode T3g(1)
was excited with the energy $E=0.2$~eV.
}
\end{figure}

Figure~\ref{fg04} shows the spatial distribution of the kinetic energy of the nonlinear localized
state generated by the excitation of the eigenmode T$_{3g}$(1). The main part of the energy
is concentrated
at the central C$_{60}$ molecule, so this state resembles discrete breathers studied earlier
for simpler lattices. The nonlinear states generated by the eigenmodes  T$_{1g}$(2) è T$_{1g}$(3)
are localized even stronger. For $E>0.01$~eV, all such modes represent strongly anharmonic
oscillations with the frequency decaying with the mode amplitude. The mode T$_{1g}$(2)
carries the largest amount of energy that can reach 1~eV.

It is important to mention that the model employing the Brenner potentials predicts
the existence of a larger number of localized modes.

In a thermalized nanocrystal all localized modes have a finite life time.
Their decay occurs mainly due to the interaction with other modes of the same molecule.
Therefore, in order to define the lifetime of a nonlinear mode, it would be sufficient
to study the dynamics of an isolated C$_{60}$ molecule with one strongly excited mode
corresponding to the nonlinear localized state and all other modes weakly thermalized.

\begin{figure}[tb]
\begin{center}
\includegraphics[angle=0, width=1\linewidth]{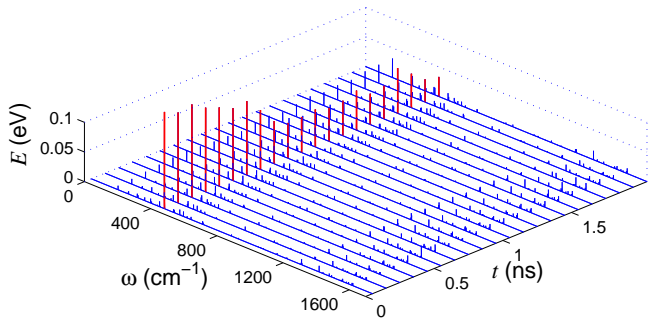}
\end{center}
\caption{\label{fg05}\protect
Interaction of a single large-amplitude mode T3g(1) with all other weakly
thermalized eigenmodes of C$_{60}$ molecules.
As the initial condition at $t=0$ we excite the eigenmode T3g(1)
with the frequency $\omega=442.3$~cm$^{-1}$ and the energy
$E_1=0.2$eV, and all other eigenmodes with the energy $E_0=0.002$eV.
Shown is the evolution of the energy distribution over the frequencies.
The peak corresponds to a large-amplitude mode (shown in red).
}
\end{figure}

In Fig.~\ref{fg05} we show the interaction of one large-amplitude eigenmode T$_{3g}$(1)
with all other weakly thermalized eigenmodes of the  fullerene C$_{60}$ molecule.
As follows from those results, the energy of the localized mode decays monotonically
exciting other modes. When the initial energy of the T$_{3g}$(1) mode is $E_1=0.2$~eV
and the energy of all other modes is $E_0=0.002$~eV, the localized mode loses a half of 
its energy for $t_0=1.4$~ns,
for $E_0=0.004$~eV --this time is $t_0=0.8$~ns,
for $E_0=0.008$~eV -- $t_0=0.4$~ns,
and for $E_0=0.016$~eV -- $t_0=0.1$~ns.
Highly-excited mode T$_{1g}$(2) with the energy  $E_1=0.2$~eV losses a half of its energy for
$t_0=0.8$~ns if energy  $E_0=0.002$~eV, for $t_0=0.5$~ns if $E_0=0.004$~eV,
for $t_0=0.3$~ns if $E_0=0.008$~eV, and for $t_0=0.1$~ns if  $E_0=0.016$~eV.
The smallest lifetime is shown by the eigenmode T$_{1g}$(3), which for the input energy
$E_1=0.2$~eV and $E_0=0.002$ looses a half of its energy for $t_0=0.1$~ns.

Therefore, our numerical results demonstrate that in a crystal formed by
C$_{60}$ molecules the spatially localized nonlinear modes are long-lived excitations,
and at room temperature their life time is of the order of several picoseconds.

\section{Thermal relaxation of C$_{60}$ crystal}

\begin{figure}[tb]
\begin{center}
\includegraphics[angle=0, width=1\linewidth]{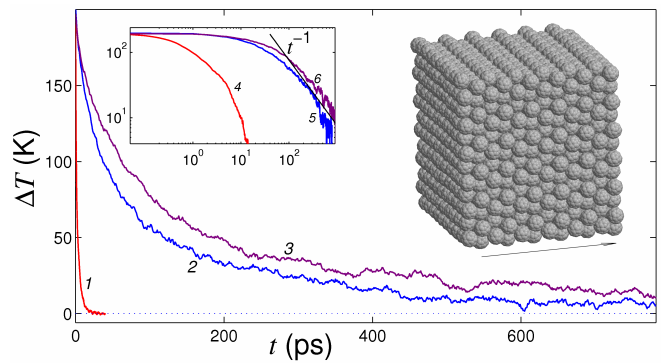}
\end{center}
\caption{\label{fg06}\protect
Relaxation the thermal gradient $\Delta T$ between the opposite sides of the
cubic cluster of $10\times 12\times 12$ fullerene C$_{60}$ molecules with time $t$
Left insert shows this dependence in the log-scale, with a straight line corresponding
to a power-law dependence, $\Delta T\sim t^{-1}$. Right insert shows a cubic nanocluster
with an arrow indicating the heat flow. Dependencies 1 and 4 correspond to the
approximation of rigid molecules when their internal dynamics is neglected, whereas
curves 2 and 5 show the results of the full molecular dynamics simulations.
Curves 3 and 6 show the results for the model employing the Brenner potentials.
}
\end{figure}

The existence of long-lived nonlinear localized collective modes in the nanocluster
of C$_{60}$ fullerene molecules should affect substantially their physical properties.
If the thermal energy can be concentrated for long time in some localized spots, such a 
crystal is expected to demonstrate low thermal conductivity. To confirm this idea, we 
consider thermal relaxation of a
cubic nanocluster composed of  $10\times12\times12$ C$_{60}$ molecules, as shown in Fig.~\ref{fg06}.

The 180-dimensional vector ${\bf u}_{nmk}$ define the coordinates of the atoms of
$(n,m,k)$-th molecule of the nanocluster ($n=1,...,10$; $m=1,...,12$; $k=1,...,12$).
For thermalizing the nanocluster, we solve the system of Langevin equations
\begin{equation}
M\ddot{\bf u}_{nmk}=-\partial H/\partial u_{nmk} -\Gamma M\dot{\bf u}_{nmk}+\Xi_{nmk},
\label{f5}
\end{equation}
where $\Gamma=1/t_r$ is the Langevin collision frequency with $t_r=0.1$~ps being
the corresponding particle relaxation time, and $\Xi_{nmk}=\{\xi_{nmk,i}\}_{i=1}^{180}$ is
a 180-dimensional vector of Gaussian distributed stochastic forces describing the interaction
of atom with the thermostat with correlation functions
$
\langle\xi_{n_1m_1k_1,i}(t_1)\xi_{n_2m_2k_2,j}(t_2)\rangle=
2M\Gamma k_BT_{n_1m_1k_1}\delta_{n_1n_2}\delta_{m_1m_2}\delta_{k_1k_2}\delta_{ij}\delta(t_2-t_1),
$
$T_{nmk}$ is the temperature of the molecule with the site $(n,m,k)$.

To obtain the initial thermalized state of the nanocluster we solve numerically the Langevin
equations (\ref{f5}) with the initial conditions corresponding to the equilibrium state. 
We run the simulations for $t=10t_r$, and then switch off the coupling with the thermostat,
i.e. neglecting the last two terms in Eqs.~(\ref{f5}), in order to study the thermal 
relaxation of the nanocluster.

We thermalize the nanocluster asymmetrically, to have only one side heated to the temperature
$T=500$~K, whereas keeping all other atoms at the temperature $T=300$K. To achieve this,
we solve the system of Langevin equations at $T_{nmk}=500$~K for the atoms with the index $k=1$
but for all other atoms with the indices $k=2,...,12$ we put $T_{nmk}=300$~K.
In  this case, we achieve a temperature different of $\Delta T=200$~K between the opposite
sides of the C$_{60}$ nanocube. As the next step, we study the thermal relaxation of the 
temperature gradient.

As a result of thermal energy exchange between the fullerene molecules, temperature difference
should decay monotonically. The crystal of fullerene molecules C$_{60}$ has a three-dimensional
structure so it should obey the Fourier law for thermal conductivity. Because time does not
enter the Fourier law, it does not allow to determine the temporal relaxation of the 
thermal gradient, for which we should employ the well-know generalization known as 
Cattaneo-Vernotte (CV) law \cite{cv1,cv2}.
In its one-dimensional version, this law can be formulated through the differential equations,
\begin{equation}
\left(1+\tau\frac{\partial}{\partial t}\right)\vec q=-\kappa\vec\nabla T,
\label{f6}
\end{equation}
where $\kappa$ is the standard coefficient of the thermal conductivity and $\tau$ is the characteristic
relaxation time of the system. In according to Eq.~(\ref{f6}), the temperature gradient 
$\Delta T$ should decay in accord with an exponential dependence.

Dependence of the temperature gradient $\Delta T$ vs. $t$ is shown in Fig.~\ref{fg06}. As follows
from this dependence, the temperature difference decays very slow, in accord with the power-law
dependence, so that it does not obey the Cattaneo-Vernotte law. In order to understand 
the reason for such an anomalous thermal relaxation as well as the role played by the 
internal dynamics of the molecules, we replace the C$_{60}$ molecules by rigid balls and 
solve the same problem assuming that the molecules in the crystal interact between each 
other as structureless particles. Our numerical results show that in this approximation 
the relaxation of the temperature difference $\Delta T$ follows an exponential dependence,
in accord with the Cattaneo-Vernotte law. Therefore, the anomalously slow relaxation
can be attributed to the excitation of long-lived nonlinear oscillations of the C$_{60}$ 
nanocluster.

\section{Conclusions}

We have applied the molecular-dynamics numerical simulations to study the nonlinear dynamics
and thermal relaxation of crystalline structures composed of C$_{60}$ fullerene molecules.
We have
revealed that such complex nanoclusters support a special type of strongly localized nonlinear
modes resembling discrete breathers of simple nonlinear lattices. For such modes,
the kinetic energy is localized primarily in the rotational modes of a single C$_{60}$ molecule
and it decays slowly inside the fullerene nanocrystal. Importantly, our results have demonstrated
that at room temperatures the lifetime of such nonlinear localized modes
may exceed tens of picoseconds. The existence of such long-lived localized nonlinear states
in the nanoclusters explains the anomalously slow thermal relaxation observed in such structures
when the temperature gradient decays in accord with the power but not exponential law, thus
violating the Cattaneo-Vernotte law of thermal conductivity.

\section*{Acknowledgements}

Alex Savin acknowledges a warm hospitality of the Nonlinear Physics Center at the Australian
National University, and he thanks the Joint Supercomputer Center of the Russian Academy 
of Sciences for the use of their computer facilities. 
The work was supported by the Australian Research Council.

\end{document}